\documentclass[aps,prb,twocolumn,showpacs,groupedaddress]{revtex4}
\usepackage{graphicx}

\begin{document}

\title{LiFeAs: An intrinsic FeAs-based superconductor with T$_c$=18 K }
\author{Joshua H. Tapp$^1$, Zhongjia Tang$^1$, Bing Lv$^1$, Kalyan Sasmal$^2$, Bernd Lorenz$^2$, Paul C.W. Chu$^{2,3,4}$, and Arnold M. Guloy$^1$}
\affiliation{$^{1}$TCSUH and Department of Chemistry, University of
Houston, Houston, TX 77204, USA} \affiliation{$^{2}$TCSUH and
Department of Physics, University of Houston, Houston, TX 77204,
USA}  \affiliation{$^{3}$Lawrence Berkeley National Laboratory, 1
Cyclotron Road, Berkeley, CA 94720, USA} \affiliation{$^{4}$Hong
Kong University of Science and Technology, Hong Kong, China}
\date{\today }

\begin{abstract}
The synthesis and properties of LiFeAs, a high-T$_c$ Fe-based
superconducting stoichiometric compound, are reported. Single
crystal X-ray studies reveal it crystallizes in the tetragonal
PbFCl-type (P4/nmm), with a = 3.7914(7) {\AA}, c = 6.364(2) {\AA}.
Unlike the known isoelectronic undoped intrinsic FeAs-compounds,
LiFeAs does not show any spin density wave behavior, but exhibits
superconductivity at ambient pressures, without chemical doping. It
exhibits a respectable transition temperature of T$_c$ = 18 K, with
electron-like carriers, and a very high critical field, H$_{c2}$(0)
$>$ 80 Tesla. LiFeAs appears to be the chemical equivalent of the
infinite layered compound of the high-T$_c$ cuprates.
\end{abstract}

 \maketitle











Until recently the chemical realm of high-T$_c$ superconductivity
had been limited mainly to copper oxide-based layered perovskites.
The latest search for noncuprate superconductors in strongly
correlated electron layered systems has led to the discovery of
high-T$_c$ superconductivity in doped quaternary rare-earth iron
oxypnictides, ROFePn, (R = rare-earth metal and Pn =
pnicogen).\cite{1,2,3} These superconductors generated enormous
interest in the materials community due to the high T$_c$'s involved
(up to 41-55 K), as well as the critical presence of a magnetic
component, Fe, considered antithetical to conventional s-wave
superconductivity.\cite{3,4} High-pressure studies suggest maximum
T$_c$ in R(O,F)FeAs may be about 50 K, but higher T$_c$'s ($>$50 K)
may yet be discovered in structurally different compounds, yet
electronically related to R(O,F)FeAs.\cite{5} Analogous alkaline
earth iron arsenides, AeFe$_2$As$_2$ (Ae = Sr, Ba), reportedly
having formal (Fe$_2$As$_2$)$^{2-}$ layers as in ROFeAs, but
separated by simple Ae-layers as in the cuprates, were found to
behave similarly.\cite{6,7} The AeFe$_2$As$_2$ phases become
superconducting (maximum T$_c\sim$37 K) with appropriate
substitution of Ae atoms with alkali metals.\cite{8,9} It was also
found that isostructural compounds, KFe$_2$As$_2$ and
CsFe$_2$As$_2$, with formal (Fe$_2$As$_2$)$^{1-}$ layers were
superconducting, with much lower T$_c$'s of 3.8 K and 2.6 K,
respectively.\cite{9} Moreover, the evolution from a superconducting
state to a spin density wave (SDW) state by chemical substitution
was observed in K$_{1-x}$Sr$_x$Fe$_2$As$_2$.\cite{9} Critical to the
high-T$_c$ FeAs-superconductors is the need to introduce sufficient
amounts of charge carriers: with electrons (n-type) by F-doping
(15-20 atm$\%$) or holes (p-type) by Sr-doping (4-13 atm$\%$) in
ROFeAs, and (K/Sr)-substitution (40:60 atm$\%$) in AeFe$_2$As$_2$.
These results established the unique role of (Fe$_2$As$_2$)-layers
in high-T$_c$ superconductivity. Since simple elemental K, Cs,
(K/Sr) or (Cs/Sr)-layers separate the (Fe$_2$As$_2$)-layers in the
AFe$_2$As$_2$ superconductors, a Li-based analog, LiFeAs, was
investigated. Its crystal structure was previously reported to be of
the Cu$_2$Sb-type that features a Fe$_2$As$_2$ substructure, similar
to the known FeAs-superconductors.\cite{10} However, the locations
of the Li atoms were problematic. Herein we report the synthesis,
single crystal structure determination, and superconducting
properties of LiFeAs, with T$_c$ = 18 K.

LiFeAs was synthesized from high temperature reactions of high
purity Li (ribbons, 99.99$\%$), Fe (pieces, 99.99$\%$) and As
(pieces, 99.99$\%$). Stoichiometric amounts of the starting
materials were placed and sealed in welded Nb tubes under Ar. The
reaction charges were jacketed within evacuated and sealed quartz
containers, and heated at 740$^o$C for 24 hours. The reaction was
then slowly cooled to room temperature over 1 day. The
polycrystalline samples of LiFeAs are black, exhibit metallic
luster, and are sensitive to moist air. All preparative
manipulations were carried out in a purified Ar-atmosphere glove
box, with total O$_2$ and H$_2$O levels $<$0.1 ppm. Elemental
analyses on single crystals and polycrystalline samples were carried
out using an inductively coupled plasma/mass spectrometer (ICPMS),
using both laser ablation and acid dissolution. Results of ICPMS
analyses indicate a Li:Fe:As mole ratio of 1.00(3):1.03(2):1.00(1),
consistent with a stoichiometric composition of LiFeAs. Phase purity
and cell parameters of the polycrystalline samples were investigated
by powder X-ray diffraction, using a Panalytical X'pert
Diffractometer. Sufficiently sized shiny single crystals of LiFeAs
were also isolated for X-ray diffraction analyses. Single crystal
X-ray structure determination was performed using a Siemens SMART
diffractometer equipped with a CCD area detector. A crystal with
dimensions of 0.28 x 0.14 x 0.02 mm$^3$ mounted in glass fiber under
a stream of cold nitrogen gas at -58$^o$C. Monochromatic Mo Ka1
radiation (=0.71073${\AA}$) was used to collect a full hemisphere of
data with the narrow-frame method. The data were integrated using
the Siemens SAINT program, and the intensities corrected for Lorentz
factor, polarization, air absorption, and absorption due to
variation in the path length. Empirical absorption correction was
applied using a lamina-shaped model, and redundant reflections were
averaged. Final unit cell parameters were refined using 512
reflections having I$>$10 $\sigma$(I).

Results of the single crystal refinement also confirm the
conclusions from the chemical analyses. LiFeAs crystallizes in a
tetragonal unit cell (P4/nmm), with a = 3.7914(7) ${\AA}$, c =
6.364(2) ${\AA}$. The major structural parameters are summarized in
Table 1. The crystal structure of LiFeAs, as shown in Figure 1, is
isostructural with the PbFCl-type, different from previous
reports.\cite{10} It is also different from the superconducting
alkali and alkaline earth metal iron arsenides, (A/Ae)Fe$_2$As$_2$
(A = K, Rb, Cs; Ae = Sr, Ba), that crystallize in the
ThCr$_2$Si$_2$-type. However, it is closely related to the "empty"
version of LaOFeAs (ZrCuSiAs-type). As in LaOFeAs, LiFeAs features
Fe$_2$As$_2$ layers, based on edge-shared tetrahedral FeAs$_4$
units. The Fe$_2$As$_2$ layers can also be derived from the
alternate As-capping of the Fe square nets, above and below each
center of the Fe squares. The Fe-As bond distance within the layers
is 2.4204(4) ${\AA}$; the nearest Fe-Fe distance is 2.6809(4)
${\AA}$. The Fe$_2$As$_2$ layers are alternately stacked, along the
c-axis, with nominal double layers of Li atoms. The parallel
stacking of the FeAs layers in LiFeAs inhibits close interlayer
contacts between As atoms. This is different from the 'slipped'
stacking in (A,Ae)Fe$_2$As$_2$ wherein adjacent Fe$_2$As$_2$ layers
are oriented by a mirror plane perpendicular to c passing through z
= 1/2 that allow closer, yet nonbonding, As-As interlayer distances.
Although the interlayer distances in LiFeAs (3.182(2) ${\AA}$) are
shorter than the ROFeAs phases, the nearest interlayer As-As
distances are long (4.2929(7) ${\AA}$). More importantly, unlike
LaOFeAs, the nominal tetrahedral sites within the nominal Li double
layers (Li-Li distances of 3.3218(4) ${\AA}$), as shown in Figure 1,
do not have any notable electron densities, and thus are unoccupied.

The results of our structure determination are different from the
conclusions of Wang et al.\cite{11} which do not agree with the
results of our work in terms of the important aspects of chemical
composition of the superconducting phase and the crystal structure.
Furthermore, we observed superconductivity in the stoichiometric
compound with an elemental ratio Li:Fe:As of 1:1:1, in contrast to
the conclusion of Ref. \cite{11} suggesting that a substantial
deficiency of Li is needed to induce superconductivity. It is
therefore important to verify the real composition which may deviate
from the nominal composition if impurity phases form during the
synthesis process. The X-ray spectra shown in Fig. 2 do not indicate
any impurity phase within the resolution of the measurement.
Furthermore, the ICPMS measurements prove the stoichiometric
composition of our sample and its uniformity within the high
resolution of the methods used. Our conclusions derived from powder
and single crystal x-ray diffraction experiments are consistent with
recent neutron powder diffraction studies.\cite{12} The slightly
lower superconducting T$_c$ of the latter work (T$_c$=16 K) is
possibly due to the off-stoichiometry (slightly Fe-richer)
composition of their samples.

\begin{table}[tbp]
\caption{Crystal structure parameters of LiFeAs. The equivalent
isotropic displacement parameter U(eq) (${\AA}^2$x10$^3$) is defined
as one third of the trace of the orthogonal U$_{ij}$ tensor. Space
group 129, cell choice 2, a=b=3.7914(7) ${\AA}$, c=6.3639(17)
${\AA}$.}
\label{1}%
\begin{ruledtabular}
\begin{tabular}{lllllll}
Atom & Wyckoff & x & y & z & U(eq) \\
As(1) & 2c & 0.25 & 0.25 & 0.2635(1) & 8(1) \\
Fe(2) & 2b & 0.75 & 0.25 & 0.50 & 7(1) \\
Li(3) & 2c & 0.25 & 0.25 & 0.8459(15) & 21(2) \\
\end{tabular}
\end{ruledtabular}
\end{table}

Magnetic susceptibility and transport measurements were performed on
single phase polycrystalline samples. The powder diffraction pattern
of the sample, as shown in Figure 2, reveals at least 22 reflections
that can be indexed to the P4/nmm space group. No impurity phase can
be resolved. Electrical resistivity as a function of temperature
$\varrho$(T) were measured using a standard 4-probe method, the
magnetic field effect on $\rho$ was determined using a Quantum
Design PPMS system for temperatures down to 1.8 K and magnetic
fields up to 7 T. The temperature dependence of the dc-magnetic
susceptibility $\chi$(T) was measured using a Quantum Design SQUID
magnetometer at fields up to 5 T. Thermoelectric power was measured
using a low frequency (0.1 Hz) ac technique with a resolution of
0.02 $\mu$V/K. During the measurements the amplitude of the
sinusoidal temperature modulation was kept constant at 0.25 K.

As shown in Figure 3, LiFeAs exhibits bulk superconductivity as
evidenced by a complete diamagnetic shielding signal, with a
superconductive transition at T$_c$=18 K. Resistivity measurements
$\varrho$T), in Figure 4, also show that LiFeAs is metallic with
$\varrho$ decreasing with temperature, and features a markedly
negative curvature. This negative curvature can be a consequence of
strong electron-electron correlation, as in KFe$_2$As$_2$, or from
strong electron-phonon interaction. The $\varrho$ dramatically drops
to almost zero below 18 K. A residual resistance observed below
T$_c$ is due to grain boundary contributions.

The superconducting transition indicated by the $\varrho$(T) and $\chi$(T) plots are consistent with the thermoelectric power S(T) data in
Figure 5. The thermoelectric power, less sensitive to grain boundaries of polycrystalline samples, is zero below T$_c$ within the resolution of
the measurement. Also, the large negative thermoelectric power of LiFeAs indicates its major carriers are electron-like (n-type), similar to the
R(O/F)FeAs superconductors, and in contrast to the hole-like carriers in KFe$_2$As$_2$ and (K/Sr)Fe$_2$As$_2$.\cite{9} Furthermore, applied
magnetic field is observed to suppress the transitions, as expected for superconducting compounds (Figure 4 inset). The critical field H vs.
T$_c$ is shown in the inset of Figure 3, with T$_c$ defined at resistivity values corresponding to changes of 5, 10, 30, and 50 $\%$ in the
total drop across T$_c$. Using the Ginzburg-Landau formula for the critical field, $H_c(t)=H_c(0)*(1-t^2)/1+t^2)$ with $t=T/T_c$, a high
zero-temperature critical field (H$_{c2}$) $>$ 80 Tesla can be extrapolated from the data derived from the 5 \% resistivity drop (Fig. 3).

LiFeAs is isoelectronic with AeFe$_2$As$_2$ and LnOFeAs, in that all
have formal (Fe$_2$As$_2$)$^{2-}$ layers. In these three compounds
elemental Li-double layers, Sr or Ba layers, or (R$_2$O$_2$) slabs
separate the charge carrying (Fe$_2$As$_2$)-layers. This arrangement
is similar to the Ae- or R-based layers that separate current
carrying (CuO$_2$)-layers in high-T$_c$ cuprates. Furthermore,
undoped parent phases of previously reported iron arsenides are not
superconducting at ambient pressures, but exhibit magnetic ordering.
Spin density wave (SDW) transitions are observed in AeFe$_2$As$_2$
(Ae = Sr, Ba)\cite{7} and LaOFeAs.\cite{1} It is therefore
surprising that intrinsic LiFeAs does not appear to show a SDW, but
exhibits superconductivity with a respectable T$_c$=18 K. The
superconducting behavior of LiFeAs can be roughly explained by
assuming incomplete charge transfer from the strongly polarizing Li
atoms to the electron-rich (Fe$_2$As$_2$)$^{2-}$ layers. However,
this would lead to hole-like behavior of the carriers, in conflict
with the thermoelectric power data. The inconsistency may arise from
changes in the conduction bands of the FeAs-layers due to variations
in inter-layer distances. Detailed band structure calculations as
recently reported\cite{13} could help to deepen the insight into the
electronic structure and the Fermi surface to understand the complex
properties of this class of compounds. High pressure experiments and
varying the electron counts on the (Fe$_2$As$_2$)$^{2\pm x}$ layers
on the superconducting properties of LiFeAs by chemical doping may
help unravel the puzzle. One may also infer that the character of
charge carriers (electrons or holes) is related to the stacking
arrangement of the (Fe$_2$As$_2$)-layers. Proper explanation of the
puzzling behavior of LiFeAs would significantly help elucidate the
mechanism of high-T$_c$ superconductivity in the layered iron
pnictides, as well as the layered cuprates. The similarities between
the layered FeAs superconductors and the layered cuprate
superconductors have been pointed out previously by us and others.
Thus, LiFeAs may be equivalent to the infinite layered member of the
high-T$_c$ cuprates.

In conclusion, a high-T$_c$ Fe-based superconductor, LiFeAs, with a
T$_c$ of 18 K and a very high H$_{c2}$ has been reported and its
structure unambiguously determined. Addition of LiFeAs to the list
of structurally different, yet isoelectronic FeAs-superconductors
provides further evidence to the important role of FeAs-layers in
this class of high-T$_c$ materials. The electron nature of the
carriers and superconductivity in undoped LiFeAs raise interesting
questions that warrant detailed band structure calculations and
further experiments.

\begin{acknowledgments}
This work was supported by the National Science Foundation
(CHE-0616805), the R.A. Welch Foundation, the United States Air
Force Office of Scientific Research, at Lawrence Berkeley National
Laboratory through the United States Department of Energy, and the
State of Texas through the Texas Center for Superconductivity.
C.W.C. also acknowledges support from the T.L.L. Temple Foundation
and the J.J. and R. Moores Endowment.
\end{acknowledgments}

\bibliographystyle{phpf}

\begin{figure}
\caption{(Color online) Crystal structure of LiFeAs with Li, Fe, and As shown as large (grey), medium (red) and small (green) spheres,
respectively. The unit cell is outlined.}

\caption{X-ray powder diffraction of LiFeAs. All peaks can be indexed with the results from the single crystal refinement. The "hump" around
16-22$^o$ is due to the X-ray scattering of the mylar film used to protect the polycrystalline samples from air.}

\caption{Magnetic susceptibility of LiFeAs. Inset: Critical fields extracted from the resisitivity data (Fig. 4) at different values (percentage
labeled) of the resistivity drop.}

\caption{(Color online) Resistivity of LiFeAs. The inset shows the field dependence near the superconducting transition.}

\caption{(Color online) Thermoelectric power of LiFeAs.}

\end{figure}

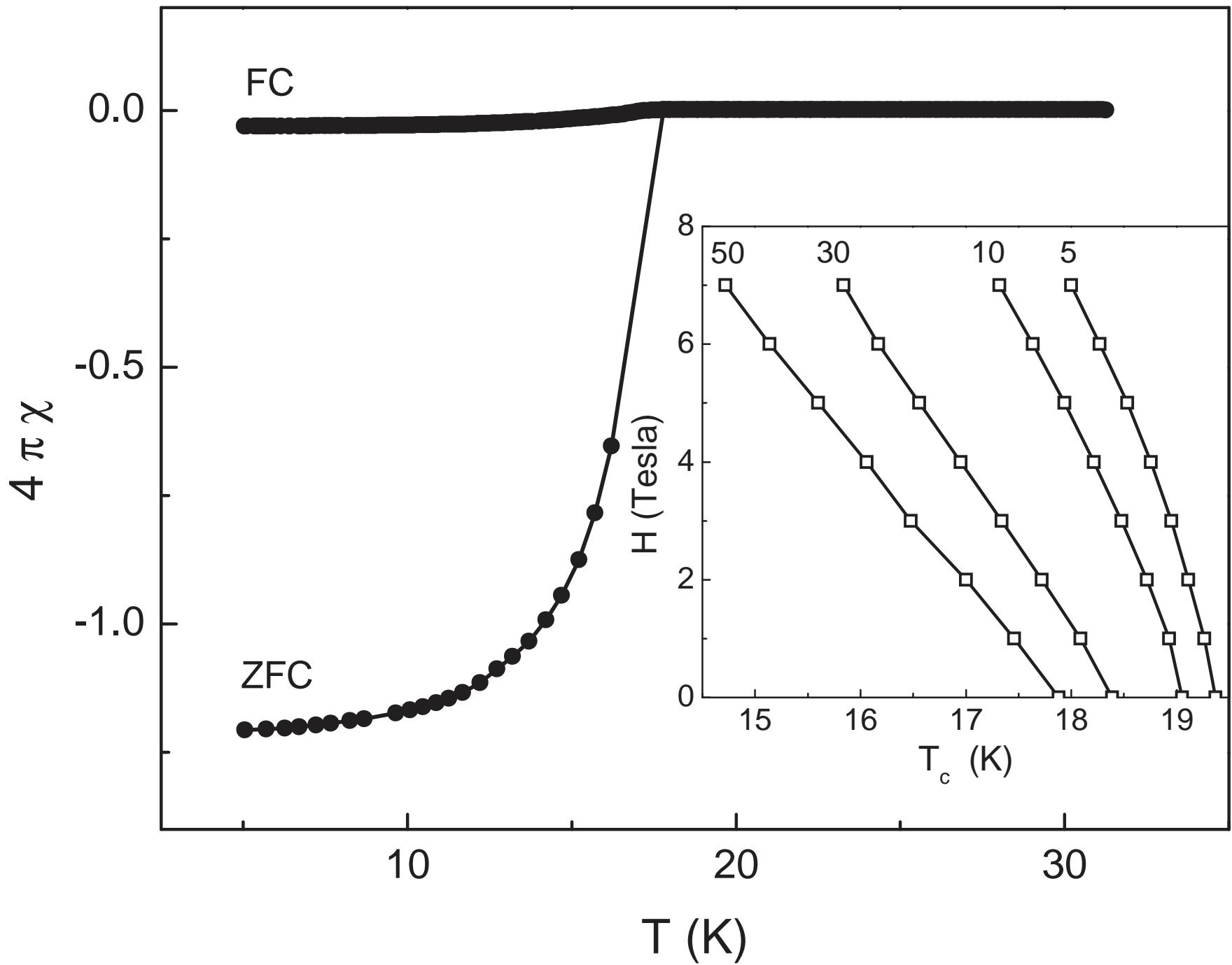
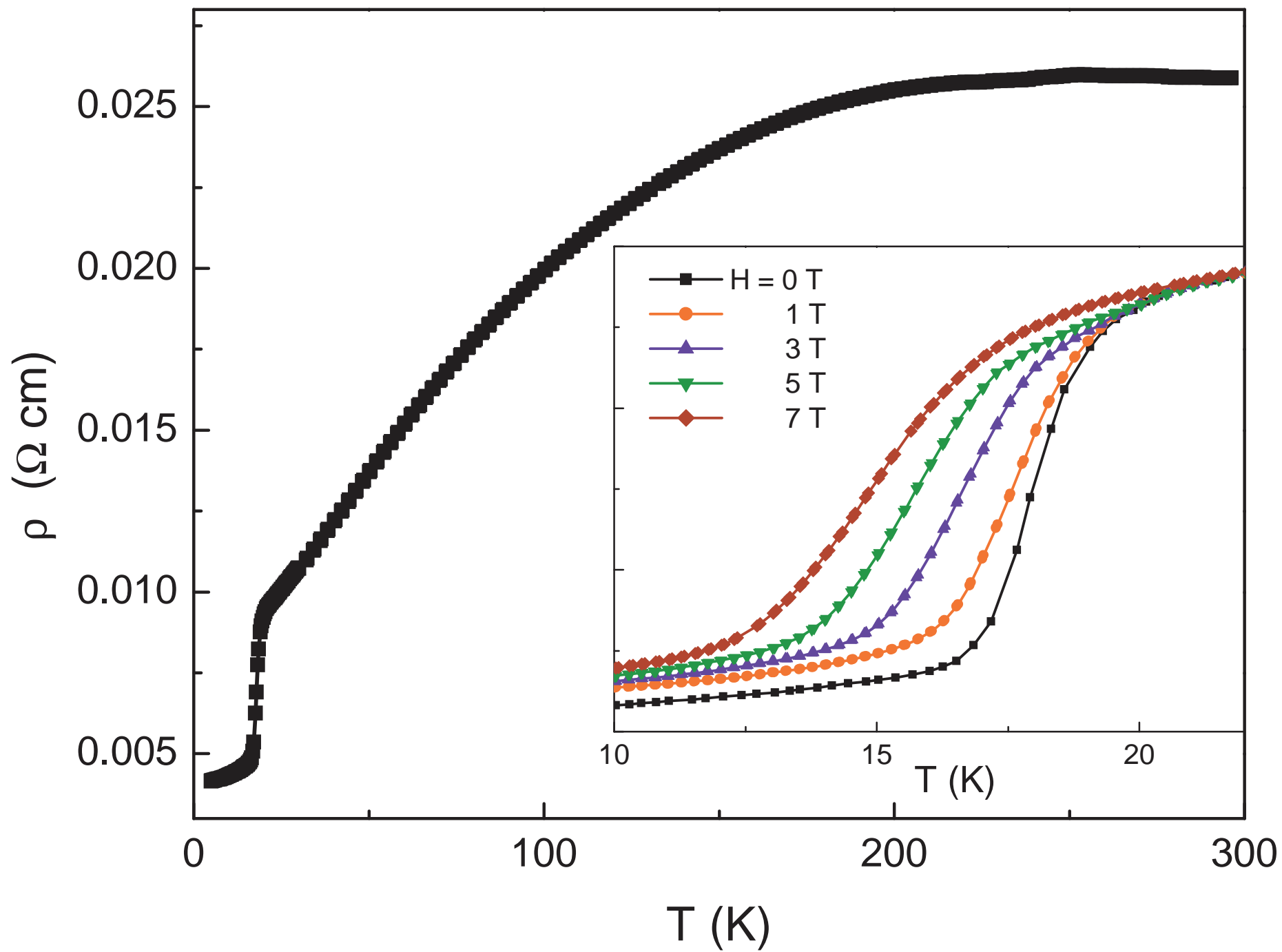
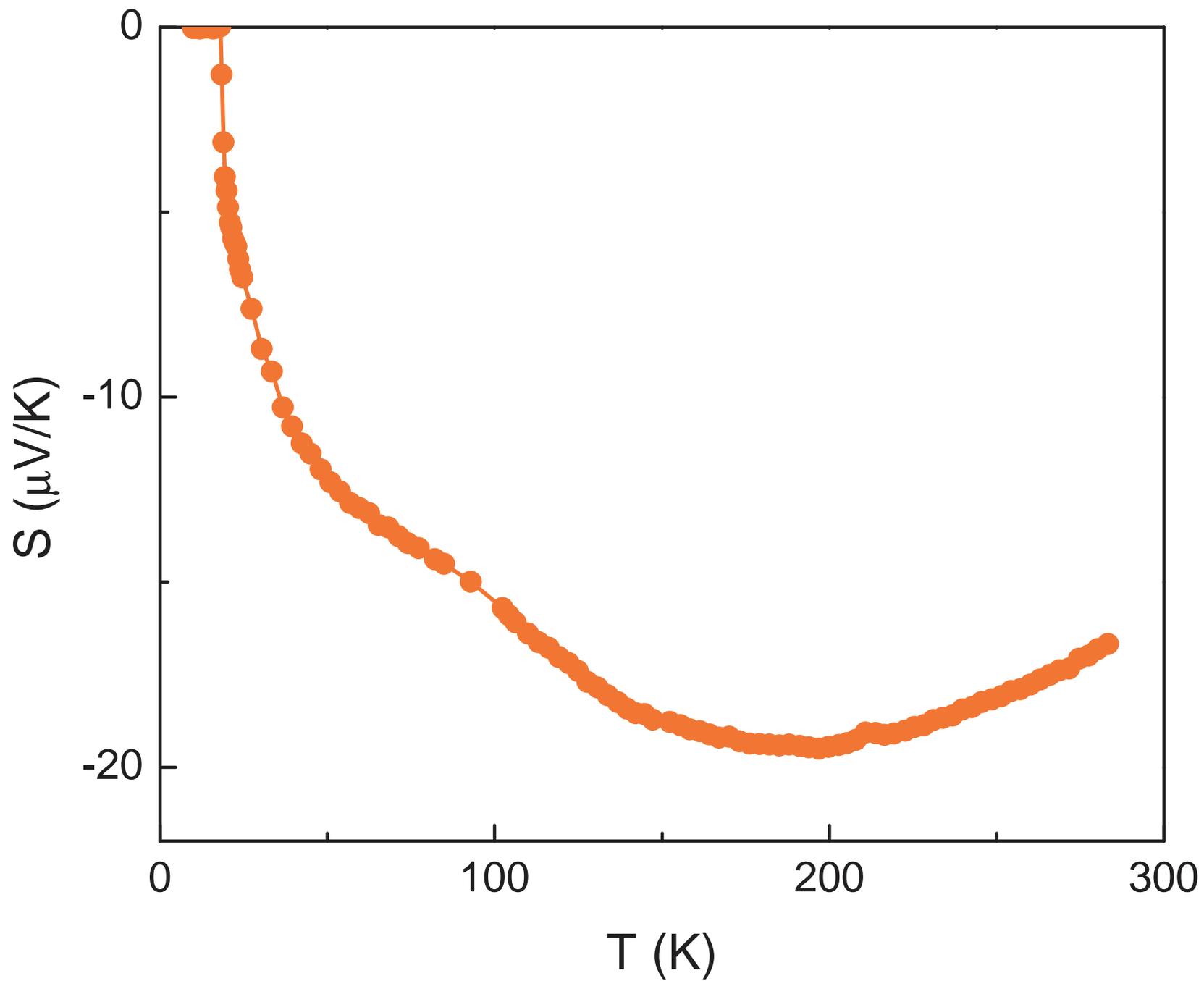
\end{document}